# Demonstration of SDN-Based Heterogeneous Quantum Key Distribution Chain Orchestration over Optical Networks

**Yuan Cao[1], Yongli Zhao[2], Jie Zhang[2], and Qin Wang[1]**

[1]*Institute of Quantum Information and Technology, Nanjing University of Posts and Telecommunications, Nanjing 210003, China*
[2]*State Key Laboratory of Information Photonics and Optical Communications, Beijing University of Posts and Telecommunications, Beijing 100876, China*

**Abstract:** Heterogeneous quantum key distribution chain orchestration over optical networks is demonstrated with software-defined networking (SDN), achieving the control-layer interoperability for BB84 (Bennett-Brassard-1984) and TF (Twin-Field) based quantum key distribution devices.

## 1. Introduction

The emergence and preliminary application of quantum key distribution (QKD) [1], [2] has opened up a new field for ensuring network information security. In practice, a central challenge limiting the widespread deployment of QKD is the scarcity and high cost of dedicated fiber resources. Deploying QKD on existing optical network infrastructures can significantly reduce the deployment difficulty and cost, while QKD can further improve the optical layer security [3]. Hence, the QKD-integrated optical network has become an innovative network paradigm, in which the telecom compatibility for QKD coexisting with high-speed optical communications has been validated in field trials [4]–[7].

Long-distance QKD between source and destination nodes usually relies on the establishment of a QKD chain [8]. Conventionally, a QKD chain is established by connecting several QKD devices running only a single-point-to-single-point QKD protocol. As an example, the Beijing-Shanghai QKD backbone [9] has been established based on the Bennett-Brassard-1984 (BB84) protocol [10]. With the continuous invention of advanced QKD protocols, the transport of secret keys end-to-end may rely on a practical QKD chain connecting multiple QKD devices running diverse QKD protocols. Such a QKD chain is referred to as the heterogeneous QKD chain. In particular, measurement-device-independent (MDI) QKD can eliminate all detection loopholes with the aid of an untrusted relay [11], while its efficient variant, namely twin-field (TF) QKD [12], can beat the PLOB bound [13] known as the secret key capacity of the point-to-point communication channel. TF-QKD achieved a long distance of 605 km in July 2021 [14]. Recently, the compatible conversion between BB84 and MDI protocols has been demonstrated in [15], [16]. Hence, a heterogeneous QKD chain with BB84 and MDI (especially TF) protocols is expected to achieve better performance than the conventional QKD chain.

Previously, the single-protocol QKD devices on each link of a QKD chain were configured independently. The complexity of configuring a QKD chain dramatically rises when it becomes

This paper is an extended version aiming to provide more details about the demonstration of heterogeneous QKD chain orchestration presented in another paper we have published elsewhere. Specifically, we have included the essential details in both papers to make them self-contained. We acknowledge the Reviewer's valuable suggestion to provide this extended version.

longer, especially in large-scale networks. Moreover, the performance (e.g., secure key rate) of QKD devices with different protocols in a heterogeneous QKD chain is difficult to be balanced. This work targets at this problem by demonstrating the software-defined networking (SDN) enabled heterogeneous QKD chain orchestration over optical networks. The SDN-based network architecture and workflow for heterogeneous QKD chain orchestration are proposed. With the aid of SDN, efficient heterogeneous QKD chain orchestration as well as the control-layer interoperability of BB84 and TF QKD devices are improved.

## 2. Network Architecture

An architecture of SDN-based QKD chain over optical networks is shown in Fig. 1, which comprises four logical layers from top to bottom, i.e., application, control, QKD chain, and optical layers. The southbound and northbound interfaces are implemented using the OpenFlow protocol and RESTful API, respectively. The southbound interface is used to exchange control and configuration request/response messages (e.g., control/configure secret-key generation and relaying), which can be realized by extending the OpenFlow Plugin and OpenFlow Java modules. The feasibility of using OpenFlow protocol in a SDN-enabled QKD network has been experimentally confirmed in [17]. The northbound interface is utilized to exchange QKD chain request/response messages, which can be realized by the POST method of Hypertext Transfer Protocol (HTTP).

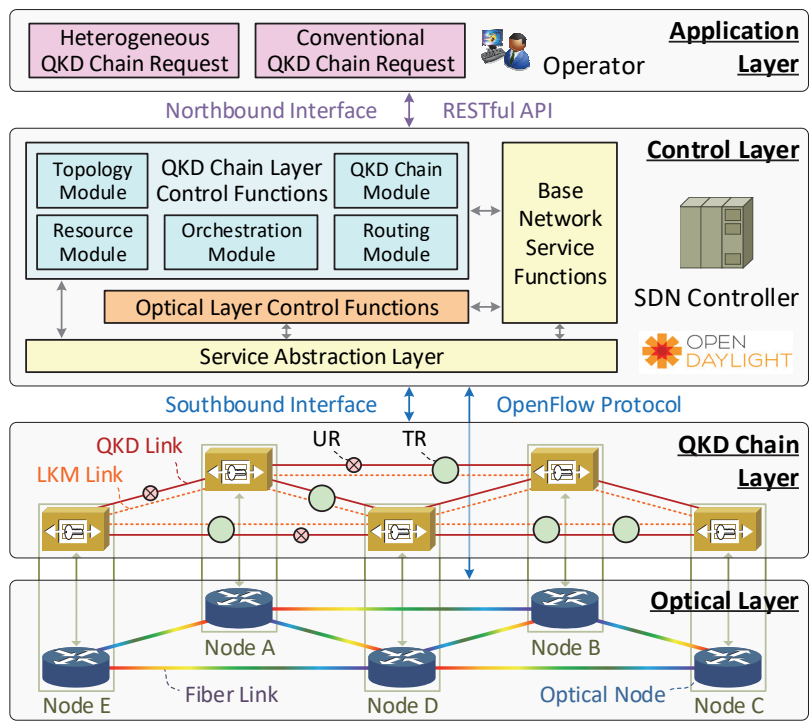

Fig. 1. An architecture of SDN-based QKD chain over optical networks.
[LKM: local key manager; UR: untrusted relay; TR: trusted relay]

**Optical Layer** consists of an existing optical network, in which a set of optical nodes are connected to each other via fiber links. Multiple services transferred over the data channel between any pair of optical nodes can be secured by QKD. Based on the wavelength-division multiplexing technique, the coexistence of the data channels for optical communications as well as the quantum and classical channels for QKD can be achieved in the same fiber [4]–[7].

**QKD Chain Layer** comprises one or more QKD chains established over the optical layer. These QKD chains can form a large-scale QKD network. The conventional and heterogeneous QKD chains are exemplified in Figs. 2(a) and 2(b), respectively, which are illustrated according to the network architecture shown in Fig. 1. A QKD chain involves two types of nodes, namely QKD nodes and relay nodes, where the relay nodes can be subdivided into trusted and untrusted relays. A QKD node is placed at the same physical location as an optical node, such that the secure keys derived from the QKD node can be delivered to the optical node for optical layer security. Each QKD chain is established between any two QKD nodes, while the relay nodes are utilized to enable the QKD chain for long reach.

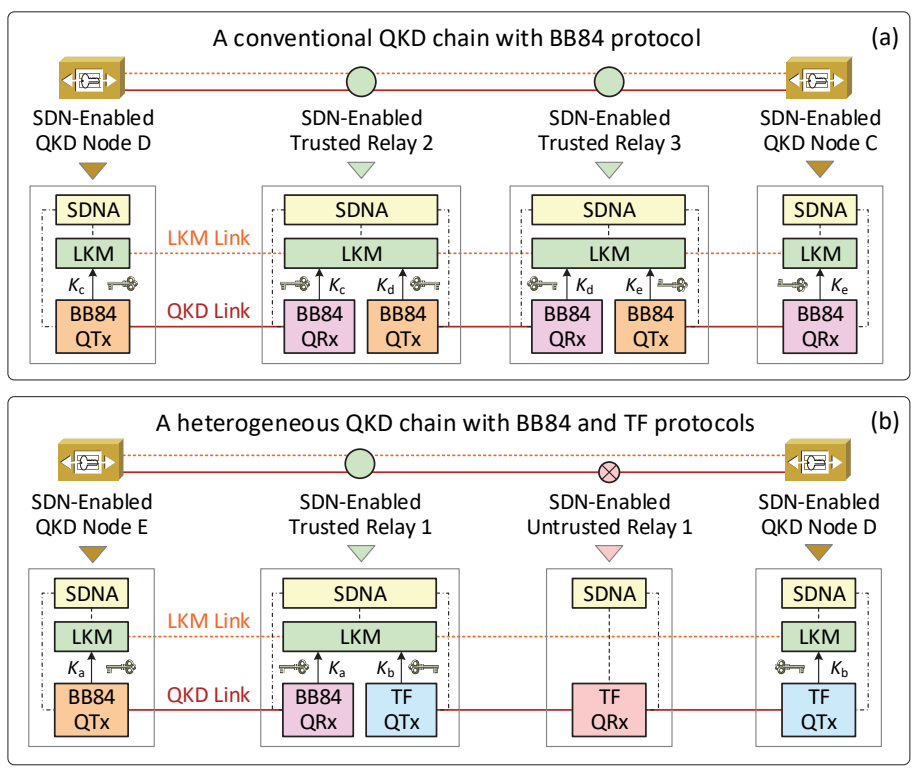

Fig. 2. Illustration of (a) a conventional QKD chain with BB84 protocol and
(b) a heterogeneous QKD chain with BB84 and TF protocols.
[SDNA: SDN agent; LKM: local key manager; QTx/QRx: QKD transmitter/receiver]

A conventional QKD chain with BB84 protocol consists of BB84 QKD transmitters (QTxs) and receivers (QRxs) connected via the QKD links, as well as local key managers (LKMs) connected via the LKM links [18]. Besides the above elements, a heterogeneous QKD chain with BB84 and TF protocols also contains TF-QTxs and TF-QRxs. Different from the BB84-QRx, the TF-QRx needs to be placed between a pair of connected TF-QTxs, which can act as an untrusted relay and remain secure even if it is accessed by an eavesdropper [12]. A QKD/relay node with an SDN agent (SDNA) is referred to as the SDN-enabled QKD/relay node. Through a short-reach interface, the SDNA can collect detailed information from its connected QKD/relay node, communicate with the SDN controller, and handle node configurations. For example, the short-reach interface can be implemented based on Internet Protocol (IP) connection between two co-located devices, whose security can be ensured by enclosing it within the security perimeter. The communications between the SDN controller and each QKD/relay node can be achieved through the protocol interpretation carried out in the SDNA. Furthermore, the relaying process performed in the heterogeneous QKD chain with BB84 and TF protocols is exemplified below.

*1) QKD Node E* encrypts the global secret key $K_G$ using $K_a$ and sends the ciphertext $K_a \oplus K_G$ to Trusted Relay 1.

*2) Trusted Relay 1* decrypts $K_G$ based on $K_a \oplus (K_a \oplus K_G) = K_G$, then encrypts $K_G$ using $K_b$ and sends the ciphertext $K_b \oplus K_G$ to QKD Node D.

*3) QKD Node D* decrypts $K_G$ based on $K_b \oplus (K_b \oplus K_G) = K_G$. Finally, the global secret key $K_G$ is shared between QKD nodes E and D.

**Control Layer** is equipped with a logically centralized SDN controller built on OpenDaylight controller platform. The control functions for QKD chain and optical layers are developed in the SDN controller. In reality, the security of SDN itself can be protected through classical and quantum countermeasures [19]. The optical layer control functions are not the focus of this work. In addition to the existing service abstraction layer and base network service functions in the OpenDaylight controller platform, five modules are designed for QKD chain layer control functions: *1) Topology Module* performs the topology abstraction as well as collects/updates the network topology and the node/link information; *2) Resource Module* stores and updates the resource (e.g., wavelength and secure key) information; *3) Orchestration Module* realizes the configuration of conventional and heterogeneous QKD chains; *4) Routing Module* computes and selects the path for QKD chain establishment; *5) QKD Chain Module* stores and updates the QKD chain information.

**Application Layer** contains the operator for multiple QKD chain requests. We consider there is only one operator over the QKD network in this work. The operator holds the information about all the conventional and heterogeneous QKD chain requests. Each QKD chain request with a specific QKD protocol requirement is generated from the operator. Specifically, a conventional QKD chain request requires the BB84 protocol, while a heterogeneous QKD chain request demands the BB84 and TF protocols in this work.

## 3. Workflow for Heterogeneous QKD Chain Orchestration

A heterogeneous QKD chain request involves source and destination QKD nodes, trusted and untrusted relays, as well as secure key rate and QKD protocol requirements. To satisfy the requirements of heterogeneous QKD chain requests triggered from the operator, the SDN controller configures the corresponding nodes via SDNAs to handle these requests one by one. The designed workflow for heterogeneous QKD chain orchestration is shown in Fig. 3. Initially, the SDN controller configures all the SDN-enabled QKD and relay nodes to accomplish QKD network initialization. After TCP session, the SDN controller achieves OpenFlow handshake and stays active with each node until the heterogeneous QKD chain request arrives, during which the detailed information of each QKD/relay node is reported to the SDN controller. The physical network topology is also obtained by the SDN controller. Furthermore, the SDN controller carries out topology abstraction and obtains the abstracted network topology. Given the untrusted relay for TF-QKD (i.e., TF-QRx) is fixed, the path across an untrusted relay between a pair of connected TF-QTxs can be predetermined. Hence, the routing computation for such a path involving three nodes (comprising two TF-QTxs and one TF-QRx, respectively) can be omitted. Subsequently, the path computation and selection for heterogeneous QKD chain requests are performed on the abstracted network topology.

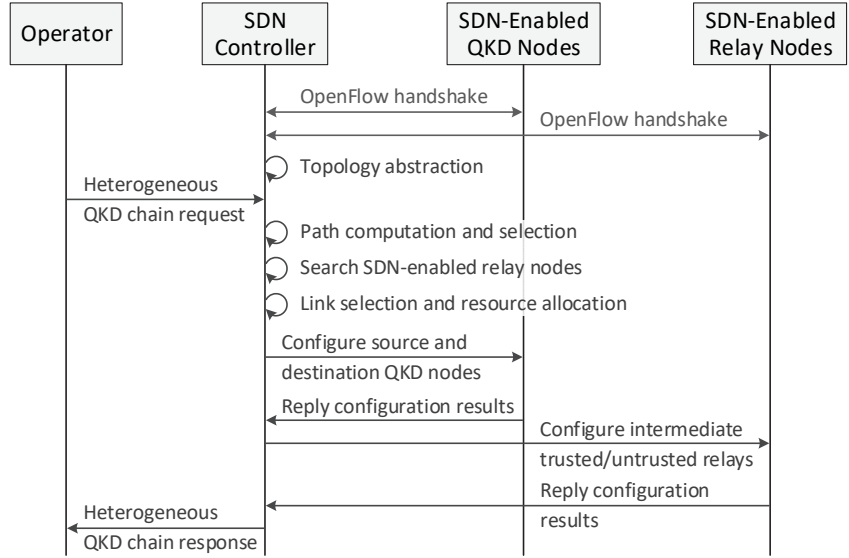

Fig. 3. Workflow for heterogeneous QKD chain orchestration.

Upon receiving a heterogeneous QKD chain request, the SDN controller first computes and selects the shortest path between source and destination QKD nodes, in which the Dijkstra's algorithm is utilized and the number of hops is considered as the routing metric. Then, the SDN controller searches the corresponding trusted and untrusted relays along the selected path. The SDN controller also performs link selection and resource (e.g., wavelength) allocation to meet the secure key rate requirement, where the first-fit algorithm [3] is employed. The first-fit algorithm is chosen in this work owing to its low complexity and small computation overhead. Using the first-fit algorithm, all the available resource units are numbered, in which a resource unit with a smaller serial number is selected and assigned before a resource unit with a larger serial number. The source and destination QKD nodes as well as the intermediate trusted and untrusted relays are configured to realize heterogeneous QKD chain orchestration. Finally, a success response is returned to the operator. The performance metrics are considered below.

*1) Total Length:* The total length of a QKD chain is characterized by the total length of QKD fiber links between source and destination QKD nodes.

*2) Secure Key Capacity (SKC):* The SKC of a QKD chain is approximately represented by the minimal secure key rate for QKD links along the selected path between source and destination QKD nodes.

*3) Control Delay:* The control delay is expressed as the time interval between the operator sending a QKD chain request and receiving a QKD chain response.

*4) Security Level:* The security level is defined as the ratio of 1 to the number of trusted relays in the QKD chain [8], since trusted relays are the weak security points in the QKD chain.

## 4. Testbed and Results

In order to support heterogeneous QKD chain orchestration, the OpenFlow protocol v1.3.0 is extended and developed to implement the southbound interface. Originally, the OpenFlow protocol was utilized to support packet switching in the electrical domain. In this study, we attach new messages to the OpenFlow protocol v1.3.0 header, which can be implemented in an OpenDaylight-based platform relying on YANG tools [20]. The OpenFlow protocol v1.3.0 header consists of the fields of version, type, length, and transaction ID. The 8-bit version field indicates the version of the OpenFlow protocol being employed. The 8-bit type field indicates the type of the extended message. In this work, the message types of heterogeneous QKD chain request and response are configured as 32 and 33, respectively. The 16-bit length field indicates the total length of the extended message. The 32-bit transaction ID field is a unique value for matching a request to the response.

Besides the OpenFlow protocol v1.3.0 header, we add 32 bits for heterogeneous QKD chain ID to represent a heterogeneous QKD chain and 32 bits for SDN-enabled QKD/relay node ID to represent an SDN-enabled QKD/relay node. Additionally, we add 32 bits (0x0000FFFF) to indicate the message function of a heterogeneous QKD chain request or response. For the heterogeneous QKD chain request, we also add 32 bits for both the input port and output port of each SDN-enabled QKD/relay node, while we add 128 bits for resource units to describe their usage. Hence, the extended message for the heterogeneous QKD chain request includes 32-bit heterogeneous QKD chain ID, 32-bit SDN-enabled QKD/relay node ID, 32-bit message function, 32-bit input port, 128-bit resource unit, and 32-bit output port. Furthermore, we add 32 bits for the heterogeneous QKD chain response to indicate the status of heterogeneous QKD chain orchestration. The extended message for the heterogeneous QKD chain response is composed of 32-bit heterogeneous QKD chain ID, 32-bit SDN-enabled QKD/relay node ID, 32-bit message function, and the status of heterogeneous QKD chain orchestration.

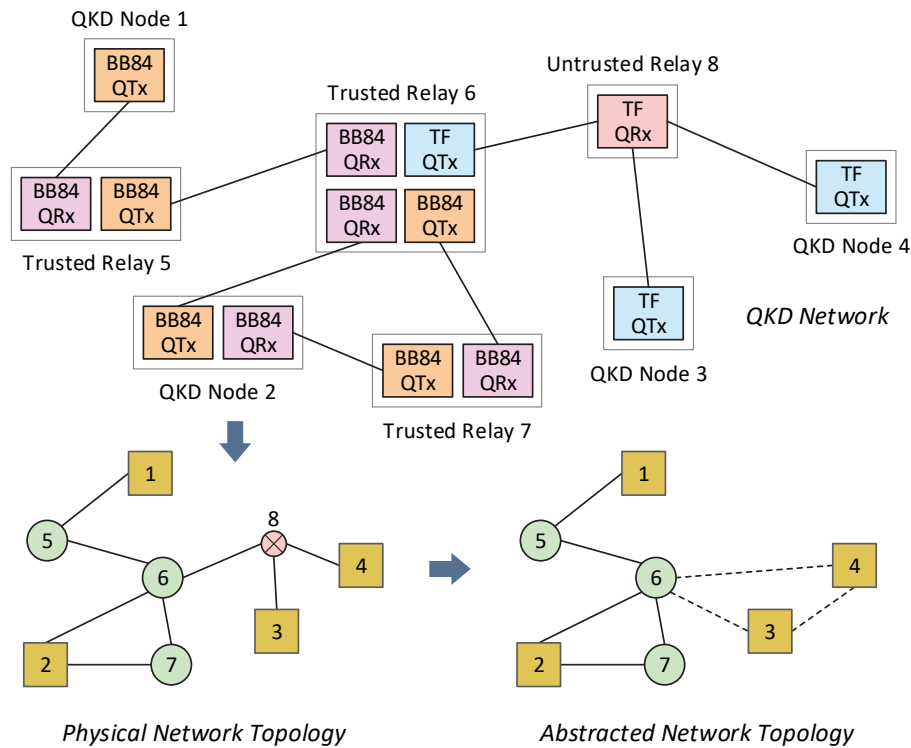

Fig. 4. Physical and abstracted network topologies corresponding to the QKD network emulated in the testbed.

The northbound interface implemented by the RESTful API is developed based on JavaScript Object Notation (JSON) to support HTTP. The heterogeneous QKD chain request/response messages between the operator and the SDN controller are implemented using the POST method of HTTP. The dedicated Uniform Resource Identifier (URI) is utilized for HTTP communication. The HTTP can be further upgraded to support HyperText Transfer Protocol Secure (HTTPS), but the cost and efficiency may be increased. Therefore, HTTPS may be more suitable for critical interfaces, such as the key delivery API specified in ETSI GS-QKD 004.

Figure 4 depicts the physical and abstracted network topologies corresponding to the QKD network emulated in the testbed, which is composed of four QKD nodes, three trusted relays, and one untrusted relay. The BB84 and TF protocols are utilized in the associated QTxs and QRxs. Based on the topology abstraction carried out by the SDN controller, the abstracted network topology is obtained, due to the fact that the path across an untrusted relay between a pair of connected TF-QTxs is predetermined. Hence, the abstracted network topology differs from the physical network topology in case of the connection of two nodes (QKD nodes or trusted relays) with an untrusted relay in between. By relying on topology abstraction, the routing computation time may be improved since the number of hops can be reduced. The Open vSwitch (OVS) is developed to function as the SDNA. Each QKD/relay node opens its control interface and data to the corresponding OVS. In this testbed, an OVS is utilized to emulate the network-layer functions of a QKD/relay node, and the detailed information (e.g., secure key rate) of each QKD/relay node is pre-stored in its connected OVS. The SDN controller is developed based on the OpenDaylight platform. The virtual machines installed in IBM servers are configured to function as the SDN controller and OVS.

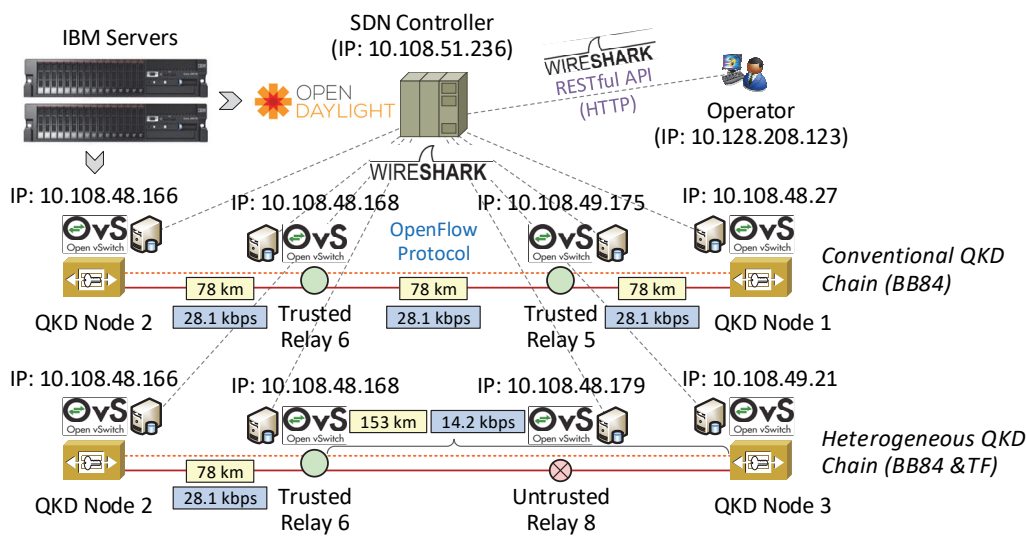

Fig. 5. Demonstration of a heterogeneous QKD chain with BB84 and TF protocols as well as a conventional QKD chain with BB84 protocol using the testbed.

For comparison purpose, a heterogeneous QKD chain with BB84 and TF protocols as well as a conventional QKD chain with BB84 protocol are demonstrated using the testbed, which are illustrated in Fig. 5. The unique IP of each SDN-enabled QKD/relay node as well as the unique IP of the operator and SDN controller are indicated. The selected path for the conventional QKD chain is along the QKD Node 2, Trusted Relay 6, Trusted Relay 5, and QKD Node 1 shown in Fig. 4, while the selected path for the heterogeneous QKD chain is along the QKD Node 2, Trusted Relay 6, Untrusted Relay 8, and QKD Node 3 shown in Fig. 4. Hence, both conventional and heterogeneous QKD chains with two relay nodes are considered in the demonstration. Each QKD link connects two adjacent QKD/relay nodes. Even without physical nodes in the emulation testbed, developing the OVS as virtual nodes can emulate the secure key generation between two connected QTx and QRx with the BB84 protocol or between two connected QTxs with the TF protocol, as well as the secure key relaying between source and destination QKD nodes. Based on the maturity of current techniques, the secure key rate for the BB84 protocol is emulated to be 28.1 kbps (corresponding to 78 km fiber link) on each point-to-point QKD link [9], while the secure key rate is emulated to be 14.2 kbps (corresponding to 153 km fiber link) between a pair of connected TF-QTxs [14].

A Wireshark network protocol analyzer is adopted to capture and analyze the HTTP and OpenFlow protocol messages. The captured results for heterogeneous QKD chain orchestration are shown in Fig. 6. The heterogeneous QKD chain request and response (HTTP messages) between the operator and SDN controller represent the start and end of the orchestration workflow, during which the SDN controller configures QKD/relay nodes via the heterogeneous QKD chain request and response (OpenFlow protocol messages). The captured OpenFlow protocol messages in Fig. 6 conform to the extended OpenFlow protocol v1.3.0 messages described above. The QKD links connecting adjacent QKD/relay nodes are orchestrated together in the heterogeneous QKD chain. Hence, the control-layer interoperability for BB84 and TF QKD devices is realized.

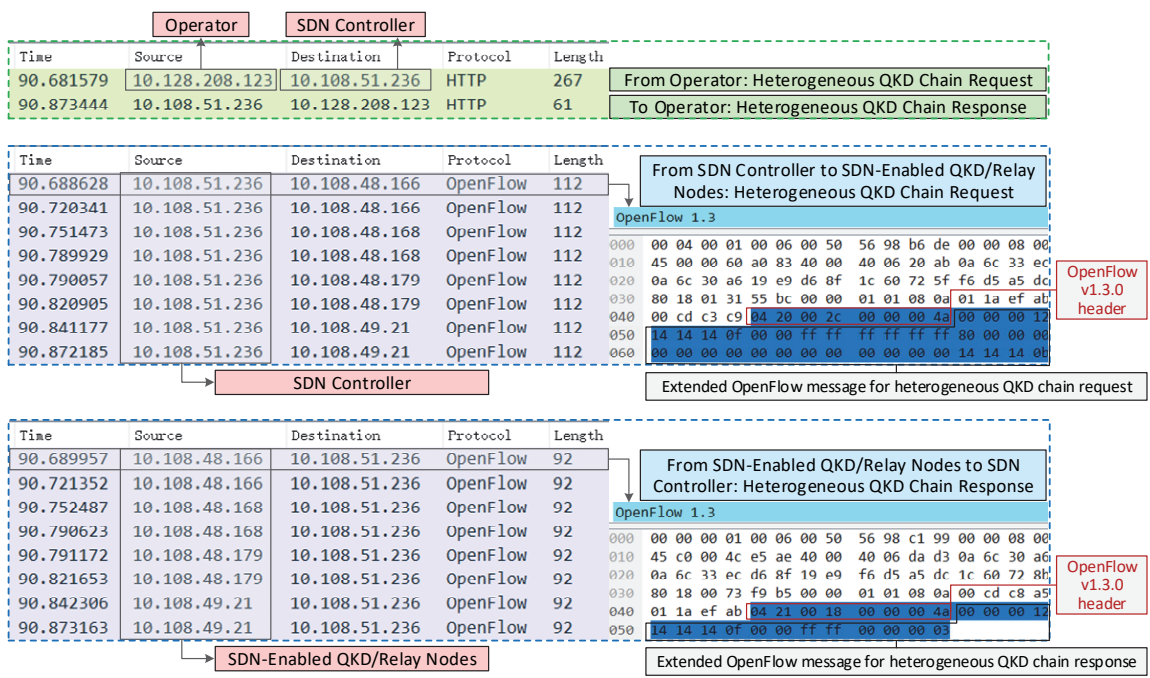

Fig. 6. The captured results (HTTP and OpenFlow protocol messages) for heterogeneous QKD chain orchestration.

The emulated results for aforementioned performance metrics are listed in Table I. The total length of the heterogeneous QKD chain is similar to that of the conventional QKD chain. The SKC of a QKD chain is mainly affected by the achievable secure key rate on each QKD link. The SKC of the heterogeneous QKD chain is lower than that of the conventional QKD chain, since the secure key rate of the TF protocol cannot match that of the BB84 protocol. In realistic environments, the BB84 protocol can provide relatively high secure key rates at the time of writing [21]. The control delay for heterogeneous QKD chain orchestration can reach 191.87 ms, which is lower than that for conventional QKD chain configuration. The reason is that each link in the conventional QKD chain is configured independently as well as the topology abstraction facilitates the routing between two nodes (QKD nodes or trusted relays) with an untrusted relay in the middle. It is important to note that the control delay in this study is software-based, while the QKD hardware latency, such as for recalibration, is typically in the range from seconds to minutes, which requires to be further demonstrated and evaluated in practical QKD networks. More details about SDN for QKD networks can be found in [4], [5], [17]. Furthermore, the heterogeneous QKD chain shows a higher level of security than the conventional QKD chain, reflecting its better security performance with BB84 and TF protocols.

Table I. Performance comparison for conventional and heterogeneous QKD chains.

| QKD Chain | Total Length | SKC | Control Delay | Security Level |
|---|---|---|---|---|
| Conventional (BB84) | 234 km | 28.1 kbps | 300.79 ms | 0.5 |
| Heterogeneous (BB84 & TF) | 231 km | 14.2 kbps | 191.87 ms | 1.0 |

## 5. Conclusion

This work achieves the demonstration of SDN-based heterogeneous QKD chain orchestration with BB84 and TF protocols in the control layer. SDN can decouple the control functions from the QKD chain layer, in which the millisecond-level control delay and high security level are realized for heterogeneous QKD chain orchestration.